\documentstyle[sprocl]{article}

\def\Journal#1#2#3#4{{#1} {\bf #2}, #3 (#4)}

\def\NPB{{\em Nucl. Phys.} B}
\def\NPA{{\em Nucl. Phys.} A}
\def\PLB{{\em Phys. Lett.}  B}

\def\PRC{{\em Phys. Rev.} C}
\def\EPJ{{\em Eur. Phys. J.} C}

\def\be{\begin{equation}}
\def\ee{\end{equation}}
\def\bea{\begin{eqnarray}}
\def\eea{\end{eqnarray}}

\begin{document}

\title{$\pi-N$ FROM AN EXTENDED EFFECTIVE FIELD THEORY}

\author{P. J. ELLIS}

\address{School of Physics and Astronomy, University of Minnesota,\\
Minneapolis, MN 55455, USA\\E-mail: ellis@physics.spa.umn.edu} 

\maketitle\abstracts{Third order chiral perturbation theory accounts for 
the $\pi-N$ scattering phase shift data out to energies slightly
below the position of the $\Delta$ resonance. The low energy constants are 
not accurately determined. Explicit inclusion of the
$\Delta$ field is favored.}

\section{Introduction}

In order to describe $\pi-N$ scattering up to at least the 
delta resonance region it would seem to be mandatory to include the
$\Delta$ isobar, along with the nucleon and pion fields, as in 
Refs.~\cite{et,fm}.
When baryon mass scales are involved in loop diagrams
the systematic counting  of chiral perturbation theory~\cite{we} in 
powers of small momenta $Q$, of order of the pion mass, is lost. 
It is restored in heavy baryon chiral perturbation theory by 
working with an effective Lagrangian in which the heavy components of the 
baryon fields have been integrated out. An alternative is to 
manipulate the loop diagrams themselves and remove the 
contributions from hard momenta of order of the baryon masses.  These 
are then absorbed in the low energy constants of the Lagrangian. 
To one loop order these two approaches give equivalent results
(the latter has been generalized in the infra-red regularization 
scheme~\cite{bl}, see Becher's talk regarding the threshold 
regime). Note that the delta-nucleon mass difference is taken to be of 
order $Q$. A practical difficulty is that the $\Delta$ propagator can go 
on shell leading to an infinite contribution to the scattering amplitude. 
In Ref.~\cite{et} this was solved by summing up the leading order, complex 
$\Delta$ self-energy diagrams so that the propagator remained finite; 
unitarity was obeyed except for a very small energy region. In Ref.~\cite{fm} 
the real part of the scattering amplitude was assumed to be the 
$K$ matrix so that unitarity was automatically obeyed.

\section{Results}

To order $Q^2$, no loops occur so it is necessary to use 
the $K$-matrix approximation. The $\pi N\Delta$ coupling can be determined 
from the $\Delta$ decay width; note that changes in the off-shell parameter 
for the vertex can be absorbed in the low energy constants~\cite{et2} so 
that any convenient value can be chosen. There are four low energy constants
to be fitted to the $S$- and $P$-wave $\pi-N$ phase shift data. 
The data can be fit~\cite{fm,eto} up to a c.m. pion kinetic 
energy of $\epsilon\sim50$ MeV (pion laboratory momentum $p\sim150$ MeV). 
Going to order $Q^3$ increases the
number of low energy constants to be fitted by about a factor of three
and a large number of loop diagrams need to be calculated.
The calculations of Refs.~\cite{et,fm} treated unitarity differently, as
mentioned, and also fitted different data sets, nevertheless common
trends appear. The fit to the phase shifts extends further to an energy 
of $\epsilon\sim100$ MeV ($p\sim250$ MeV), slightly below the 
$\Delta$ resonance position. These results show a considerable improvement 
over an order $Q^3$ calculation~\cite{fms} which excluded the $\Delta$ field, 
particularly in the $P33$ channel as might be expected. They are comparable 
to, perhaps a little better than, an order $Q^4$ calculation~\cite{fmo} 
with just the $\pi$ and $N$ fields. It is also worth remarking that the 
convergence of the chiral series is much better when the $\Delta$ field is 
explicitly included. At order $Q^3$ the $\pi NN$ vertex is dressed by a 
series of loop diagrams, nevertheless the effective coupling remains close
to the tree-level value~\cite{et} and the Goldberger-Treiman discrepancy is 
a few percent in agreement with a determination from pionic hydrogen data 
and the GMO sum rule~\cite{sc}. The effective $\pi N\Delta$ couping also 
remains close to the tree-level value. The low energy constants are not 
accurately 
determined and unconstrained fits often produce unreasonable values for the 
sigma term. Nevertheless the fits are only slightly degraded if it is 
constrained. A value close to the canonical~\cite{gls} 45 MeV is 
obtained~\cite{fm} with the Karlsruhe data, whereas a larger figure 
$\sim75$ MeV is favored~\cite{et} with the Virginia Polytechnic
Institute data.

\section*{Acknowledgments}
I acknowledge partial support from the US Department of Energy under
grant no. DE-FG02-87ER40328.

\end{document}